\documentclass{article}[20.5pt]
\usepackage[margin=1in]{geometry}
\usepackage[utf8]{inputenc}
\usepackage[natbibapa]{apacite}
\usepackage{amsmath,amssymb,array}
\usepackage{booktabs}
\usepackage[lined,boxed,linesnumbered]{algorithm2e}
\usepackage{hyperref}
\usepackage{comment}
\usepackage{graphicx}
\usepackage{moodle}
\RequirePackage{fancyvrb}
\RequirePackage{alltt}
\usepackage{etex}
\DefineVerbatimEnvironment{example}{Verbatim}{}
\renewenvironment{example*}{\begin{alltt}}{\end{alltt}}
\usepackage{authblk}

\input{preamble}

\title{PDXpower: A Power Analysis Tool for Experimental Design in Pre-clinical Xenograft Studies for Uncensored and Censored Outcomes}

\author[1,2]{Shanpeng Li}
\author[2]{Donatello Telesca}
\author[3,4]{Harley I. Kornblum}
\author[5]{David Nathanson}
\author[6,7]{Frank Pajonk}
\author[2]{Elvis Han Cui}
\author[1]{Joycelynne Palmer}
\author[2]{Gang Li}

\affil[1]{Division of Biostatistics, Department of Computational and Quantitative Medicine, City of Hope, Duarte, USA.}
\affil[2]{Department of Biostatistics, University of California, Los Angeles, USA.}
\affil[3]{Departments of Psychiatry, Pharmacology and Pediatrics, University of California at Los Angeles, Los Angeles, CA, USA.}
\affil[4]{The Intellectual and Developmental Disabilities Research Center, University of California, Los Angeles, USA.}
\affil[5]{Department of Molecular and Medical Pharmacology, University of California, Los Angeles, USA.}
\affil[6]{Department of Radiation Oncology, University of California, Los Angeles, USA.}
\affil[7]{Department of Neurosurgery, University of California, Los Angeles, USA.}

\begin{document}

\maketitle

%\title{\thanksref{t1}}
%\thankstext[id=t1]{}

\begin{abstract}
In cancer research, leveraging patient-derived xenografts (PDXs) in pre-clinical experiments is a crucial approach for assessing innovative therapeutic strategies. Addressing the inherent variability in treatment response among and within individual PDX lines is essential. However, the current literature lacks a user-friendly statistical power analysis tool capable of concurrently determining the required number of PDX lines and animals per line per treatment group in this context.
In this paper, we present a simulation-based R package for sample size determination, named `\textbf{PDXpower}', which is publicly available at The Comprehensive R Archive Network \url{https://CRAN.R-project.org/package=PDXpower}. The package is designed to estimate the necessary number of both PDX lines and animals per line per treatment group for the design of a PDX experiment, whether for an uncensored outcome, or a censored time-to-event outcome. 
Our sample size considerations rely on 
two widely used analytical frameworks: the mixed effects ANOVA model for uncensored outcomes and Cox's frailty model for censored data outcomes, which effectively account for both inter-PDX variability and intra-PDX correlation in treatment response.
Step-by-step illustrations for utilizing the developed package are provided, catering to scenarios with or without preliminary data. 

\end{abstract}

\section{Introduction}

Xenograft studies involve the transplantation of cells, tissues, or organs from one species, known as the donor species, into another species, called the recipient or host species. In these studies, researchers typically implant human cells, tissues, or tumors into animal models, commonly rodents, to investigate various aspects of human biology, disease progression, and therapeutic interventions. In the context of cancer research, pre-clinical experiments through patient-derived xenografts (PDXs) provide an important scientific tool for the evaluation of novel therapeutic strategies. In particular, PDX studies take into account the high level of response variability between subjects through subject-specific derived replication across treatment groups.

%A high level of between-subject variability is indeed commonly identified in clinical research, and power calculations targeting population level response heterogeneity can be challenging.  Our primary focus in this manuscript is to provide a framework for sample size and power calculations in pre-clinical PDX studies, with the specific goal of concurrently choosing both the number of PDX lines (subjects) and the number of animals (repeated measures) per PDX line. 

One of the experimental strengths of pre-clinical PDX studies stems from the investigator's ability to observe the results of a quasi-counterfactual treatment assignment protocol, which sees the same patient-derived tumor potentially treated under different conditions through replication across genetically homogeneous animal models.   
\citet{eckel2021experimental} discussed a taxonomy of PDX designs encompassing {\it nested}, {\it crossed}, and {\it mixed crossed/nested} designs, as depicted in Figure 2 of their paper. The {\it nested} design involves a naïve hierarchy in which different PDX lines are randomized between treatment groups, before PDX-specific replication is obtained within group and PDX line. While seemingly reasonable, this design does not fully address the possibility of tumor-specific replicates assigned to different treatment groups, and remains susceptible to high levels of tumor-specific response heterogeneity.   
At the opposite end, a {\it crossed} design avoids confounding of randomization and PDX lines by assessing how each level of a factor impacts the outcome with all other factors, i.e. testing all treatments on the same animal grown within each of the PDX lines. As a theoretical construct, this design creates multiple PDX lines and administers multiple drugs within the same animal concurrently, but it often fails to yield meaningful results and lacks feasibility, thus becoming impractical in most experimental settings.
Alternatively, incorporating elements from both protocols, 
a mixed crossed/nested design 
allows every PDX line to be evaluated for all
treatments (crossed design), while allowing subsampling of
PDX using animals that cannot be reused across treatment
groups (nested design). Effectively addressing potential confounding between PDX
line and treatment group, the mixed crossed/nested design has become a common workhorse in PDX research. 
Using retrospective data from IDH-wildtype glioblastoma preclinical experiments evaluating three treatments across 27 PDX lines,
\citet{eckel2021experimental} demonstrated through empirical simulations that experimental designs employing few animals across many PDX lines can yield robust results and accommodate inter-tumor variability.

The purpose of this paper is to introduce a new statistical \textbf{R} package, named \textbf{PDXpower}, designed for power analysis and determination of the required number of PDX lines and animals per line per treatment group in a PDX experiment structured under the {\it mixed crossed/nested} design framework. Notably, statistical power and sample size considerations for testing a treatment effect depend on various factors, including experiment design, statistical model, effect size, and prior information derived from either preliminary data or previous studies. As detailed in Section \ref{sec:method}, for uncensored survival time, we employ the mixed effects ANOVA model \citep{rosner2015fundamentals} using PDX as a random effect, which accounts for both intra- and inter-PDX variability, naturally applicable to the mixed crossed/nested design. While several statistical power analysis methods and software packages exist for mixed effects ANOVA models across different analytical platforms, including \textbf{R} \citep{R, clusterPower, simr, longpower, pamm, clusterPower, lu2008sample, Liu1997, kumle2021estimating}, \textbf{SAS} \citep{SAS}, and \textbf{PASS} \citep{PASS}, determining the required number of PDX lines and animals per line per treatment group concurrently often necessitates additional coding effort, potentially complex or time-consuming for those less familiar with coding. Our developed \textbf{R} package, \textbf{PDXpower}, addresses this gap, automating the process of determining both parameters simultaneously for the mixed effects ANOVA model. 
In additional to power analysis based on the mixed effects ANOVA model for uncensored data, our package also includes a module for power calculation with right-censored time-to-event data using Cox's frailty model \citep{duchateau2008frailty,rondeau2012frailtypack}. Particularly, this module is tailored for Type 1 censoring of a survival outcome with a fixed administrative censoring time for all animals, which is typical for pre-clinical animal studies. Similar to the mixed effects ANOVA model, the existing power analysis tools for Cox's frailty models are currently limited to determining the required number of PDX lines with a pre-specified number of animals per line per treatment group \citep{chen2014sample,dinart2024sample}. To the best of our knowledge, our R package is the first accessible statistical tool to simultaneously determine both the required number of PDX lines and the number of animals per line per treatment group.

The rest of the paper is organized as follows. In Section \ref{sec:method}, we specify the mixed effects ANOVA model for testing treatment effects with uncensored data, and Cox's frailty model for right-censored data. We also evaluate the Type 1 error rate and rejection power of the statistical test across varying number of PDX lines and animals per line per treatment group though simulations. In Section \ref{sec:example}, we provide a tutorial on utilizing our developed package \textbf{PDXpower} to conduct power analysis for different scenarios, with or without preliminary data. Additional remarks are provided in Section \ref{sec:dis}.

\section{Sample size determination}
\label{sec:method}
\subsection{Notation and Statistical Formalization of the Mixed Crossed/Nested Design}
\label{sect:2.1}
For ease of exposition and without loss of generality, we consider comparing outcomes between the control and treatment group, i.e., group $A$ and group $B$. In the mixed crossed/nested design, for each of $n$ PDX patients/cell lines, we have $2\times m$ implanted animals: $m$ animals are  randomized to treatment $A$ and the other $m$ animals are randomized to treatment $B$. 
Let $i \in \{1,2,\ldots,n\}$ represent the index of the PDX lines,  $j \in \{1,2,\ldots,2m\}$ denote the index of the animals within each PDX line, and $D_{ij}$ be a treatment indicator, where $D_{ij} = 1$ if animal $j$ within PDX line $i$ is in group $B$, and $0$ otherwise. We denote by $Y_{ij}$ the outcome of interest, such as the time to death since the beginning of treatment, for animals $j$ within PDX line $i$,  $i=1,\ldots, n$ and $j=1,\ldots, 2m$. The observed data will consist of $\{(Y_{ij}, D_{ij}): i=1,\ldots,n; j=1,\ldots,2m\}$.

\subsection{Statistical models for mixed crossed/nested design}
We consider two popular analytical frameworks applicable to the mixed crossed/nested design discussed earlier: a mixed effects ANOVA model for uncensored data and Cox's frailty model for right-censored data. 

\begin{enumerate}
    \item  Mixed effects ANOVA model: For $i=1,\ldots,n; j=1,\ldots,2m$,
    \begin{eqnarray}
    \label{ANOVA}
    \log Y_{ij} = \beta_0 + D_{ij}\beta + \alpha_{i} + \epsilon_{ij},
    \end{eqnarray}
    where $\beta_0$ is the intercept, $\beta$ is the treatment effect, $\alpha_i \sim_{iid} N(0,\tau^2)$ represents an unobserved PDX-specific random effect, and $\epsilon_{ij} \sim_{iid} N(0,\sigma^2)$ is a random residual error specific to animals within all PDX lines.
    %Based on the assumptions of independence of $\alpha_i$ and $\epsilon_{ijk}$, $\log Y_{ijk}$ are also assumed normally distributed and independent given $\alpha_i$.
    \item Cox's frailty model: For $i=1,\ldots,n; j=1,\ldots,2m$,
    \begin{eqnarray}
    \label{Cox}
    \lambda_{ij}(t) = \lambda_0(t|\lambda, \nu)\exp\{D_{ij}\beta + \alpha_i\},
    \end{eqnarray}
    where $\lambda_{ij}(t)$ is the hazard function of mouse $j$ within PDX $i$, $\lambda_0(t|\lambda, \nu)$ is a baseline hazard function, following a 2-parameter Weibull distribution $Weibull(\lambda, \nu)$, with scale and shape parameters $\lambda$ and $\nu$, respectively, $\beta$ represents the treatment effect, and $\alpha_i\sim_{iid} N(0,\tau^2)$ is an unobserved PDX-specific random effect (frailty). As in Model 1, given $\alpha_i$, $Y_{ijk}$ are assumed to be mutually independent. %The key assumption of this model is the proportional hazards assumption, i.e., the hazard ratio is constant over time within the same PDX line (conditional on $\alpha_i$). 
\end{enumerate}

Under each of the above models, a Wald-type test can then be conducted for the null hypothesis $H_0: \beta=0$ to assess the treatment effects at a pre-specified significance level. 

\subsection{Simulations}
\label{sect:simulation}
We present some simulations to evaluate and illustrate the performance of the Wald test under the mixed effects ANOVA model for uncensored data and Cox's frailty model for right-censored data for varying number of PDX lines 
$n $ and 
animals per line per treatment group $m$.
%are reported for all possible combinations based on 500 Monte Carlo replicates.

\subsubsection{Simulation 1: Mixed effects ANOVA model for uncensored data}
\label{sec:simulation1}
\begin{figure}
    \centering
    \includegraphics[width=9cm]{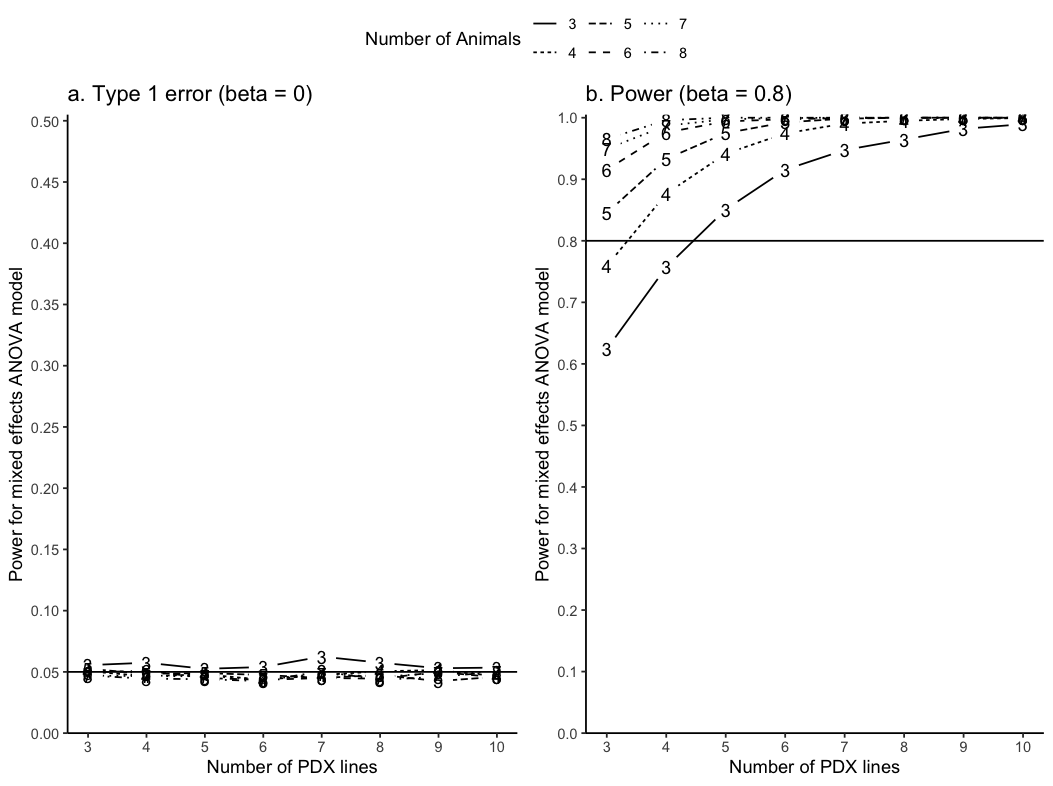}
    \caption{Estimated rejection power of a level $\alpha=0.05$ Wald test for $\beta=0$ using the mixed effects ANOVA model (\ref{ANOVA}) based on 2,000 Monte Carlo samples generated from model (\ref{simANOVA}) for  various combinations of PDX lines 
$n = (3,4,5,6,7,8,9,10)$ and 
animals per line per treatment group  $m = (3, 4, 5, 6, 7, 8)$ under two scenarios: (a)  $\beta=0$ (left panel), and (b) $\beta=0.8$ (right panel).}
    \label{fig:FigANOVAAP}
\end{figure}

We considered two treatment effect scenarios: (a)  $\beta=0$, and (b) $\beta=0.8$. For each treatment effect scenario and  combination of PDX lines 
$n = (3,4,5,6,7,8,9,10)$ and 
animals per line per treatment group $m = (3, 4, 5, 6, 7, 8)$, we generated 2000 Monte Carlo samples according to the mixed crossed/nested design described in Section \ref{sect:2.1} and the following   mixed effects ANOVA model:
    \begin{eqnarray}
\label{simANOVA}
\log Y_{ij} &=& 5 + D_{ij}\beta + \alpha_{i} + \epsilon_{ij}, \\
\alpha_{i} &\sim& N(0, 0.2), \nonumber \\
\epsilon_{ij} &\sim& N(0, 0.5). \nonumber
\end{eqnarray}
The estimated rejection power, defined as the proportion of instances where the null hypothesis $H_0:\beta=0$ is rejected at $\alpha =5\%$ significance level using the mixed effects ANOVA model (\ref{ANOVA}), is summarized in Figure \ref{fig:FigANOVAAP} for each treatment effect scenario and  combination of PDX lines 
$n $ and 
animals per line per treatment group $m $. 

From Figure \ref{fig:FigANOVAAP}a (left panel), it is evident that the Monte Carlo estimates of the type I error rates closely align with the nominal level $\alpha = 0.05$ across all combinations of $n$ and $m$. Additionally, Figure \ref{fig:FigANOVAAP}b (right panel) illustrates that increasing the number of PDX lines or animals per line enhances the statistical power to detect the treatment effect, as expected.
%, which is consistent with the conclusions in \citet{eckel2021experimental}. 

\subsubsection{Simulation 2: Cox's frailty model for censored data}
\label{sec:simulation2}
Similar to Simulation 1, we also considered
two treatment effect scenarios: (a)  $\beta=0$, and (b) $\beta=0.8$. For each treatment effect scenario and  combination of PDX lines 
$n = (3,4,5,6,7,8,9,10)$ and 
animals per line per treatment group  $m = (3, 4, 5, 6, 7, 8)$, we generated 2000 right-censored Monte Carlo samples according to the mixed crossed/nested design described in Section \ref{sect:2.1} and the following Cox's frailty model:
%\item Censored data: Generate $Y_{ijk}$ from Cox's frailty model: 
\begin{eqnarray}
\label{simCox}
{Y}_{ijk} &\sim& 0.3\exp(D_{k}\beta + \alpha_i), \nonumber\\
\alpha_{i} &\sim& N(0, 0.2), \nonumber 
%\lambda_0(t|\lambda, \nu) &\sim& \text{exponential}(0.3), \nonumber\\
%Y_{ijk} &=& \text{min}(\tilde{Y}_{ijk}, C),
\end{eqnarray}
where $Y_{ijk}$ is subject to right-censoring at a pre-determined time point $C=8$ (end of follow-up).
%where $\lambda=0.3, \nu = 1, C = 8$. Following this mechanism (\ref{simCox}), $Y_{ijk}$ is observed to be a possibly censored time-to-event outcome, subject to the type I censoring mechanism $C$, a pre-determined fixed time point.

\begin{figure}
    \centering
    \includegraphics[width=9cm]{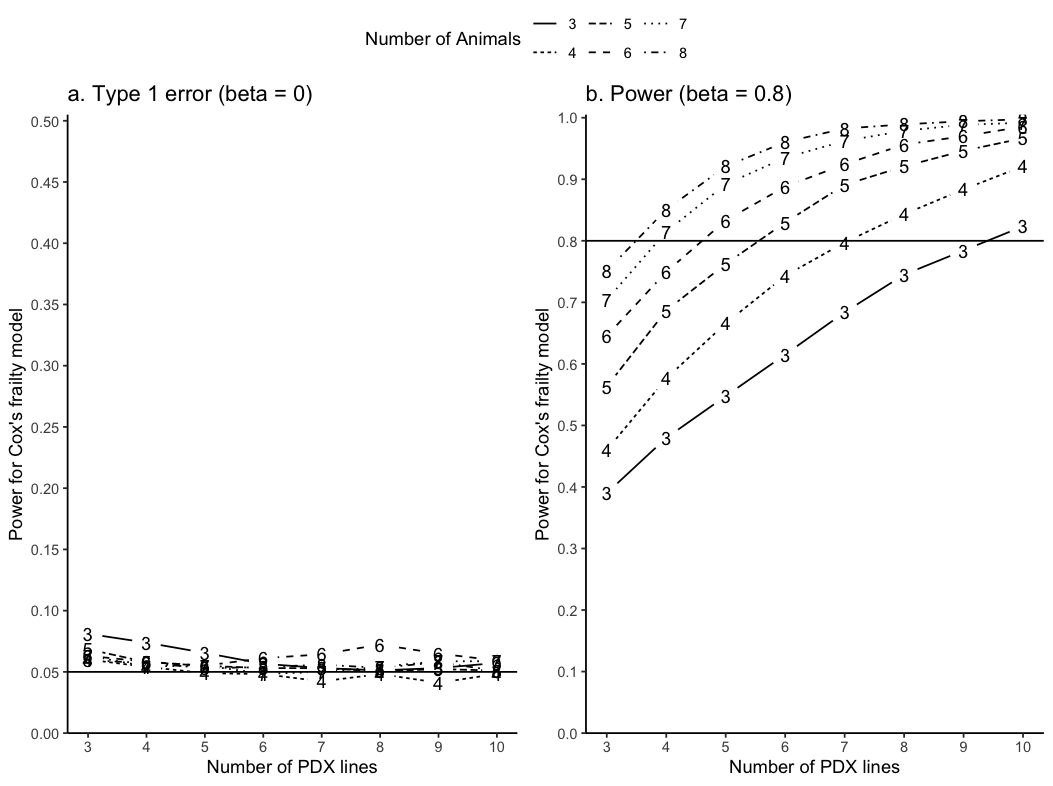}
    \caption{Estimated rejection power of a level $\alpha=0.05$ Wald test for $\beta=0$ using Cox's frailty  model (\ref{Cox}) based on 2000 Monte Carlo samples generated from model (\ref{simCox}) for  various combinations of PDX lines 
$n = (3,4,5,6,7,8,9,10)$ and 
animals per line per treatment group  $m = (3, 4, 5, 6, 7, 8)$ under two scenarios: (a)  $\beta=0$ (left panel), and (b) $\beta=0.8$ (right panel).
}
    \label{fig:FigCoxAP}
\end{figure}

 The estimated rejection power of the Wald-type test for testing $\beta=0$ at a significance level of $\alpha =5\%$ based on Cox's frailty model (\ref{Cox}) is depicted in Figure \ref{fig:FigCoxAP} for two treatment effect scenarios and various combinations of $n$ and $m$. It is observed that the Monte Carlo estimates of the type I error rates  closely match the nominal level $\alpha = 0.05$ across all combinations of $n$ and $m$ (Figure \ref{fig:FigCoxAP}a), and increasing the number of PDX lines or animals per line per treatment group enhances the statistical power to detect the treatment effect (Figure \ref{fig:FigCoxAP}b).
 % This is consistent with the conclusions of  \citet{eckel2021experimental} who considered fixed effects models. 

\subsection{Statistical power calculation and sample size determination via simulation}
\label{subsect3.2}
We now outline a simulation-based strategy to obtain Monte Carlo estimate of the statistical power for assessing a treatment effect using the mixed crossed/nested design in PDX animal experiments. This strategy is based on the mixed effects ANOVA model (\ref{ANOVA}) for uncensored data and Cox's frailty model (\ref{Cox}) for right-censored data, as discussed in Sections \ref{sec:ANOVApower} and \ref{sec:COXpower}, respectively. 

\subsubsection{Simulation-based power calculation for the mixed crossed/nested design with uncensored data}
\label{sec:ANOVApower}
To perform simulation-based power calculation for the mixed crossed/nested design with uncensored data using the mixed effects ANOVA model, the following information must be determined a priori:
\begin{enumerate}
    \item[i.] The scientific hypothesis: $H_a:\beta=\beta_1$, where $\beta_1$= anticipated difference in median log(survival time) between the treatment and control groups.
    
    \item[ii.] Statistical significance level $\alpha$. It is a commonly used threshold to control the Type 1 error rate, the probability of concluding the results are statistically significant when, in reality, they were arrived at purely by chance. 
    
    \item[iii.] Sample sizes $n$ and $m$. $n$ is the number of PDX lines and $m$ is the number of animals per PDX line per treat group.
    
    \item[iv.] Error variance $\sigma^2$ in the mixed effects ANOVA model (\ref{ANOVA}). This parameter quantifies the unexplained variation across PDX lines and animals.
    
    \item[v.] Inter-PDX variability $\tau^2$. It quantifies the inter-PDX variation across PDX lines. 
    
    \item[vi.] The number of Monte Carlo replicates \texttt{sim}. 
    
\end{enumerate}

With the above priori information, we employ the following simulation strategy for power calculation and subsequently determine the number of PDX lines $n$ and number of animals $m$ per PDX line per group within the mixed effects ANOVA model framework for uncensored data.
\begin{enumerate}
    \item Define a range of feasible values for $n$ and a range of feasible values for $m$.  Subsequently, for every combination of $n$ and $m$, follow the subsequent steps.
    \item Generate a Monte Carlo sample $\{(Y_{ij},D_{ij}), \ i=1,\ldots, n; j=1,\ldots, 2m\}$ 
    according to the mixed crossed/nested design described in Section \ref{sect:2.1} and the mixed effects ANOVA model 
    (\ref{ANOVA}) with priori information i-vi.
    \item Fit the mixed effects ANOVA model (\ref{ANOVA}) on the simulated data and test $H_0:\beta = 0$ at significance level $\alpha$.
    %In Model 1, the outcome is defined as the natural logarithm of the time to death or time to euthanization, with censored animals excluded from the analysis. In Model 2, the outcome is defined as the minimum of time to death or time to euthanization and censoring time. 
    %Treatment effect is estimated and tested for a significant difference between two treatment groups using a two-sided test. Effect sizes and p-values are obtained for both models for each experimental design. A threshold of $\alpha=0.05$ is considered statistically significant.
    \item Repeat steps 2 and 3 over \texttt{sim}  Monte Carlo samples. Calculate the estimated power as the proportion of instances where the null hypothesis $H_0:\beta = 0$ is rejected.
    %\item Plot the estimated power for all combinations of $n$ and $m$ as exemplified  in Figure \ref{Fig4:Weibull2}.
    \item Given a desired power, say 80\%, determine the minimal required number of PDX lines and number of animals per line per treatment group by examining the estimated power across all combinations of $n$ and $m$.% from Figure \ref{Fig4:Weibull2}.   
\end{enumerate}

\subsubsection{Simulation-based power calculation for the mixed crossed/nested design with right-censored data}
\label{sec:COXpower}
To conduct simulation-based power calculation for the mixed crossed/nested design with right-censored data using Cox's frailty model (\ref{Cox}), the following priori information are required.
\begin{enumerate}
    \item[i.] The scientific hypothesis: $H_a:\beta=\beta_1$, where $e^{\beta_1}$ represents the hazard ratio between the treatment and control groups. 
    \item[ii.] Statistical significance level $\alpha$. %It is a commonly used threshold to control the Type 1 error rate, the probability of concluding the results are statistically significant when, in reality, they were arrived at purely by chance. 
    \item[iii.] Sample sizes $n$ and $m$. $n$ is the number of PDX lines and $m$ is the number of animals per PDX line per treat group.
    
    \item[iv.] Inter-PDX variability $\tau^2$. It is  the variance of PDX-specific random effect $\alpha_i\sim N(0, \tau^2)$, which  quantifies the inter-PDX variation across PDX lines.
    \item[v.] Baseline hazard in Cox's frailty model (\ref{Cox}). Most existing power analysis tools for survival data assume the exponential or Weibull distribution as a working model for the baseline hazard, which is also adopted in our simulation-based power calculation. 
    \item[vi.] Duration of follow-up. In pre-clinical studies, it is typical to initiate treatment for all animals simultaneously and follow them until the conclusion of the study, resulting in type I censoring of the time-to-event outcome at the end of the follow-up period.
    %This censoring mechanism is also available in our power analysis tool if there is a need to determine the total study period in a simulation. 
    \item[vii.] The number of Monte Carlo replicates \texttt{sim}. 
    %Typically, at least 500 Monte Carlo replicates are necessary to obtain stable estimates of power.
\end{enumerate}

With the above priori information, the following simulation strategy will be used for power calculation and subsequently determine the number of PDX lines $n$ and number of animals $m$ per PDX line within Cox's frailty model framework for right-censored data.
\begin{enumerate}
    \item Define a range of feasible values for $n$ and a range of feasible values for $m$.  Subsequently, for every pairing of $n$ and $m$, follow the subsequent steps.
    \item Generate a Monte Carlo sample $\{(Y_{ij},D_{ij}), \ i=1,\ldots, n; j=1,\ldots, 2m\}$ according to the mixed crossed/nested design described in Section \ref{sect:2.1} and Cox's frailty model (\ref{Cox}) with the above priori information i-vii. Then, form a right-censored sample $\{(\tilde{Y}_{ij}, \delta_i, D_{ij}) \equiv (\min\{Y_{ij},C\}, I(Y_{ijk}\le C), D_{ij}): \ i=1,\ldots, n; j=1,\ldots, 2m\}$.
    \item Fit Cox's frailty model (\ref{Cox}) on the simulated data and test $H_0:\beta = 0$ at significance level $\alpha$.
    %In Model 1, the outcome is defined as the natural logarithm of the time to death or time to euthanization, with censored animals excluded from the analysis. In Model 2, the outcome is defined as the minimum of time to death or time to euthanization and censoring time. 
    %Treatment effect is estimated and tested for a significant difference between two treatment groups using a two-sided test. Effect sizes and p-values are obtained for both models for each experimental design. A threshold of $\alpha=0.05$ is considered statistically significant.
    \item Repeat steps 2 and 3 over \texttt{sim} Monte Carlo samples. Calculate the estimated power as the proportion of instances where the null hypothesis $H_0:\beta = 0$ is rejected.
    %\item Plot the estimated power for all combinations of $n$ and $m$ as exemplified  in Figure \ref{Fig4:Weibull2}.
    \item Given a desired power, say 80\%, determine the minimal required number of PDX lines and number of animals per line per treatment group by examining the estimated power across all combinations of $n$ and $m$.% from Figure \ref{Fig4:Weibull2}.   
\end{enumerate}

\section{A Hands-On Tutorial of \textbf{PDXpower}}
\label{sec:example}
We have created an R package, \textbf{PDXpower}, to implement the simulation-based power analysis strategy outlined in the previous section. This package functions as a user-friendly analytical tool for determining the required number of PDX lines and animals per line per treatment group under the mixed crossed/nested design for preclinical PDX experiments. Below, we offer practical guidelines and illustrative examples, demonstrating the utilization of \textbf{PDXpower} for designing preclinical PDX experiments for both uncensored and censored data.

\subsection{Power analysis for the mixed crossed/nested design with uncensored data}
\label{sec:uncensoredexample}

\subsubsection{Power analysis based on preliminary data}
\label{sec:pilotdata}
In certain pre-clinical studies, researchers may possess preliminary data from initial or related experimental explorations before embarking on a larger study. Priori information, as discussed in Section \ref{sec:ANOVApower},  can be derived from the preliminary data, aiding  simulation-based power analysis to determine the required number of PDX lines and animals per line per treatment group.

First, install and load the package in a local RStudio environment by running the following code:
\begin{example}
install.packages("PDXpower")
require(PDXpower)
\end{example}

For illustration purpose, we have generated an uncensored preliminary dataset named \verb'animals1' through simulation, which is stored in our package. The \verb'animals1' dataset comprises 18 animals and includes three columns: ID (PDX line ID number), Y (survival time of each animal), and Tx (treatment indicator with 1 for treatment and 0 for control). A screenshot of this dataset, generated as an output of the following R code, is displayed below. 
\begin{example}
## load the dataset from the package
data(animals1)
animals1

   ID          Y Tx
1   1  0.8059523  0
2   1  0.6847214  0
3   1  1.4467016  0
4   1  1.4520716  1
5   1  1.6975660  1
6   1  0.8451278  1
7   2  0.8291675  0
8   2  1.1890589  0
9   2  1.6846127  0
10  2  3.0847914  1
11  2  2.4491112  1
12  2  5.0783589  1
13  3  1.0568824  0
14  3  0.7999763  0
15  3  1.6059887  0
16  3  0.9706344  1
17  3 11.7938743  1
18  3  1.4031069  1
\end{example}
Next, a call to the function \texttt{PowANOVADat} first fits the mixed effects ANOVA model (\ref{ANOVA}) on \verb'animals1'. Subsequently, using the resulting parameter estimates as prior information, it conducts simulations to obtain Monte Carlo estimates of the power function across various sample size combinations.
\begin{example}
### Power analysis by assuming the time to event 
### is log-normal distributed
PowTab <- PowANOVADat(data = animals1, 
formula = log(Y) ~ Tx, 
random = ~ 1|ID, 
n = 3:10, m = 2:8, sim = 500)
\end{example}
The inputs for the function \texttt{PowANOVADat} include \texttt{data}, \texttt{formula}, the number of PDX lines \texttt{n}, the number of animals \texttt{m}, and the number of Monte Carlo replicates \texttt{sim}. The output, saved in PowTab, is shown below.

\begin{example}
Parameter estimates based on the pilot data:
Treatment effect (beta): 0.7299 
Variance of random effect (tau2): 0.0332 
Random error variance (sigma2): 0.386 

Monte Carlo power estimate, calculated as the
  proportion of instances where the null hypothesis
  H_0: beta = 0 is rejected (n = number of PDX lines,
  m = number of animals per arm per PDX line,
  N = total number of animals for a given combination 
  of n and m):
    n m   N Power (%)
1   3 2  12      49.6
2   3 3  18      67.0
3   3 4  24      77.4
4   3 5  30      87.2
5   3 6  36      94.2
6   3 7  42      96.0
7   3 8  48      98.2
8   4 2  16      64.6
9   4 3  24      79.6
10  4 4  32      88.0
11  4 5  40      96.0
12  4 6  48      98.2
13  4 7  56      99.4
14  4 8  64      99.8
15  5 2  20      72.6
16  5 3  30      86.8
17  5 4  40      94.8
18  5 5  50      98.6
19  5 6  60      99.2
20  5 7  70      99.8
21  5 8  80     100.0
22  6 2  24      80.4
23  6 3  36      92.8
24  6 4  48      97.4
25  6 5  60      99.8
26  6 6  72     100.0
27  6 7  84     100.0
28  6 8  96     100.0
29  7 2  28      85.6
30  7 3  42      96.2
31  7 4  56      99.4
32  7 5  70     100.0
33  7 6  84     100.0
34  7 7  98     100.0
35  7 8 112     100.0
36  8 2  32      89.0
37  8 3  48      98.4
38  8 4  64     100.0
39  8 5  80     100.0
40  8 6  96     100.0
41  8 7 112     100.0
42  8 8 128     100.0
43  9 2  36      92.2
44  9 3  54      98.8
45  9 4  72      99.8
46  9 5  90     100.0
47  9 6 108     100.0
48  9 7 126     100.0
49  9 8 144     100.0
50 10 2  40      95.8
51 10 3  60      99.4
52 10 4  80      99.8
53 10 5 100     100.0
54 10 6 120     100.0
55 10 7 140     100.0
56 10 8 160     100.0


\end{example}

\begin{figure}
    \centering
    \includegraphics[width=9cm]{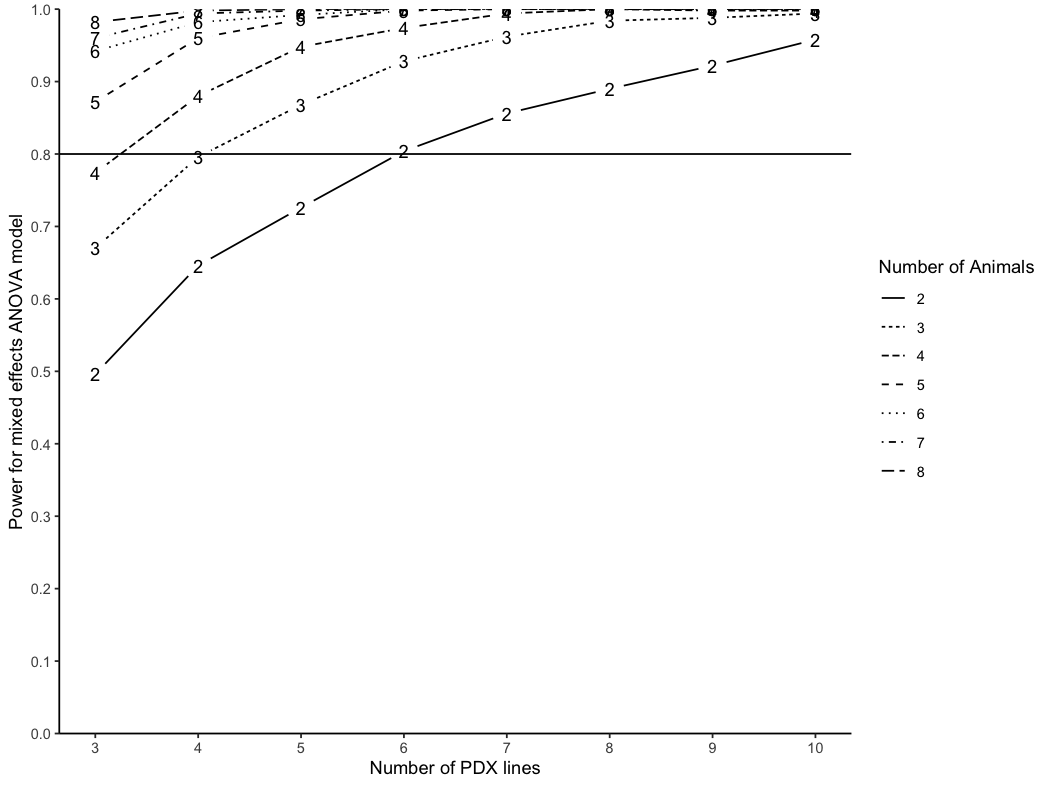}
    \caption{Power curve of the illustrating example for the mixed effects ANOVA model with preliminary data.}
    \label{fig:FigPower2}
\end{figure}

One can also visualize the results from the above table, as shown in Figure \ref{fig:FigPower2}, by calling the function \texttt{plotpower} in our package \textbf{PDXpower}. 
\begin{example}
### Visulaize the above table
plotpower(PowTab[[4]], ylim = c(0, 1))
\end{example}
 where \texttt{ylim} specifies the range of the y-axis in the plot. 
%Figure \ref{fig:FigPower2} shows the trajectory of estimated power for different sample size combinations.

It is observed from Figure \ref{fig:FigPower2} that to achieve approximately 80\% power at a 5\% significance level, the minimal sample size of PDX lines and animals per arm per line per treatment group can be any of the following $(n,m)$ combinations: $\{(3, 4), (4, 3), (6, 2)\}$. An investigator can then decide which combination to choose based on the feasibility of the study. For example, if it is feasible to employ 6 PDX lines, one may prefer to choose the (6, 2) combination to ensure the maximum possible information on PDX line heterogeneity, resulting in a total of 24 animals (6 PDX lines * 2 treatments * 2 animals per line per treatment group).

\subsubsection{Power analysis without preliminary data}
\label{sec:nopilotdata}
When preliminary data on the treatment under investigation are not available, one may conduct power analysis using relevant information from previous or related studies to estimate the necessary priori information, sometimes incorporating additional model assumptions. Table \ref{tab:tab1} provides a summary of the required information for power calculations using the mixed effects ANOVA model for the mixed crossed/nested PDX experimental design in the absence of preliminary data. 

\begin{table}
\resizebox{9cm}{!} 
{ 
\begin{tabular}{@{}llll}
\hline \hline
Parameter & Interpretation   & Values\\
\hline
\\[-.05in]
{\bf ANOVA}\\
\\[-.1in]
ctl.med.surv & Median survival in the control arm & Any positive number\\
tx.med.surv &Median survival in the treatment arm & Any positive number\\[.1in]
$\sigma^2$ & Error variance      &  Any positive number (set $\sigma^2 = 1$ as default) &\\
      %     & $-$  Set to assumed variance of  & set $\sigma^2 = 1$ as default\\
        %   & log survival distribution &\\[.1in]
icc  & Intra-PDX correlation coefficient  & 0 $<$ icc $<$ 1 (set icc = 0.1 as default)\\
 %         &    & icc $\rightarrow$ 0: lower heterogeneity\\
 %          & & icc $\rightarrow$ 1: higher heterogeneity\\
\hline
 \hline
\end{tabular}
}
\caption{Parameter value elicitation for Log-Normal distributed data for given number of PDX lines $n$ and animals per cell line $m$.}
\label{tab:tab1}
\end{table}

If one assumes the time-to-event outcome follows a log-normal distribution, then a power analysis will be performed using the mixed effects ANOVA model. The underlying model assumptions require specifying a median survival time in the control and treatment groups, an error variance ($\sigma^2$, set 1 as default) for the log-survival distribution, and the intra-PDX correlation coefficient (icc, set 0.10 as default), quantifying the proportion of PDX heterogeneity over the total variation, i.e., $\tau^2 / (\tau^2 + \sigma^2)$. 
For example, assuming a median survival of 2.4 time units (days, months, etc.) in the control group, 7.2 time units in the treatment group, and an intra-PDX correlation coefficient icc=0.10, a call to the function \texttt{PowANOVA} generates log-normal Monte Carlo samples (see equation (\ref{simANOVA})) and conducts Monte Carlo simulation-based power calculation over sample size combinations, as shown below:
{\small
\begin{example}
### Power analysis by specifying the median survival
### of control and treatment group and assuming 
### the time-to-event is log-normal distributed
PowTab <- PowANOVA(ctl.med.surv = 2.4,
tx.med.surv = 7.2, icc = 0.1, sigma2 = 1, sim = 500,
n = 3:10, m = 2:8)
\end{example}
}
The output, saved in PowTab, is shown below.
{\small
\begin{example}
Treatment effect (beta): -1.098612 
Variance of random effect (tau2): 0.1111111 
Intra-PDX correlation coefficient (icc): 0.1
Random error variance (sigma2): 1 

Monte Carlo power estimate, calculated as the
  proportion of instances where the null hypothesis
  H_0: beta = 0 is rejected (n = number of PDX lines,
  m = number of animals per arm per PDX line,
  N = total number of animals for a given combination 
  of n and m):
    n m   N Power (%)
1   3 2  12      43.0
2   3 3  18      61.0
3   3 4  24      78.0
4   3 5  30      82.8
5   3 6  36      89.8
6   3 7  42      93.2
7   3 8  48      97.2
8   4 2  16      54.8
9   4 3  24      74.6
10  4 4  32      88.8
11  4 5  40      93.2
12  4 6  48      96.4
13  4 7  56      98.6
14  4 8  64      98.8
15  5 2  20      67.0
16  5 3  30      83.2
17  5 4  40      93.2
18  5 5  50      96.6
19  5 6  60      98.4
20  5 7  70      99.8
21  5 8  80      99.6
22  6 2  24      74.0
23  6 3  36      92.0
24  6 4  48      97.6
25  6 5  60      99.2
26  6 6  72      99.0
27  6 7  84     100.0
28  6 8  96      99.8
29  7 2  28      81.8
30  7 3  42      94.4
31  7 4  56      98.6
32  7 5  70      99.4
33  7 6  84      99.8
34  7 7  98      99.8
35  7 8 112     100.0
36  8 2  32      87.2
37  8 3  48      96.8
38  8 4  64      99.4
39  8 5  80      99.6
40  8 6  96      99.8
41  8 7 112     100.0
42  8 8 128     100.0
43  9 2  36      89.2
44  9 3  54      98.0
45  9 4  72      99.6
46  9 5  90     100.0
47  9 6 108     100.0
48  9 7 126     100.0
49  9 8 144     100.0
50 10 2  40      92.8
51 10 3  60      99.4
52 10 4  80      99.6
53 10 5 100     100.0
54 10 6 120     100.0
55 10 7 140     100.0
56 10 8 160     100.0
\end{example}
}

\begin{figure}
    \centering
    \includegraphics[width=9cm]{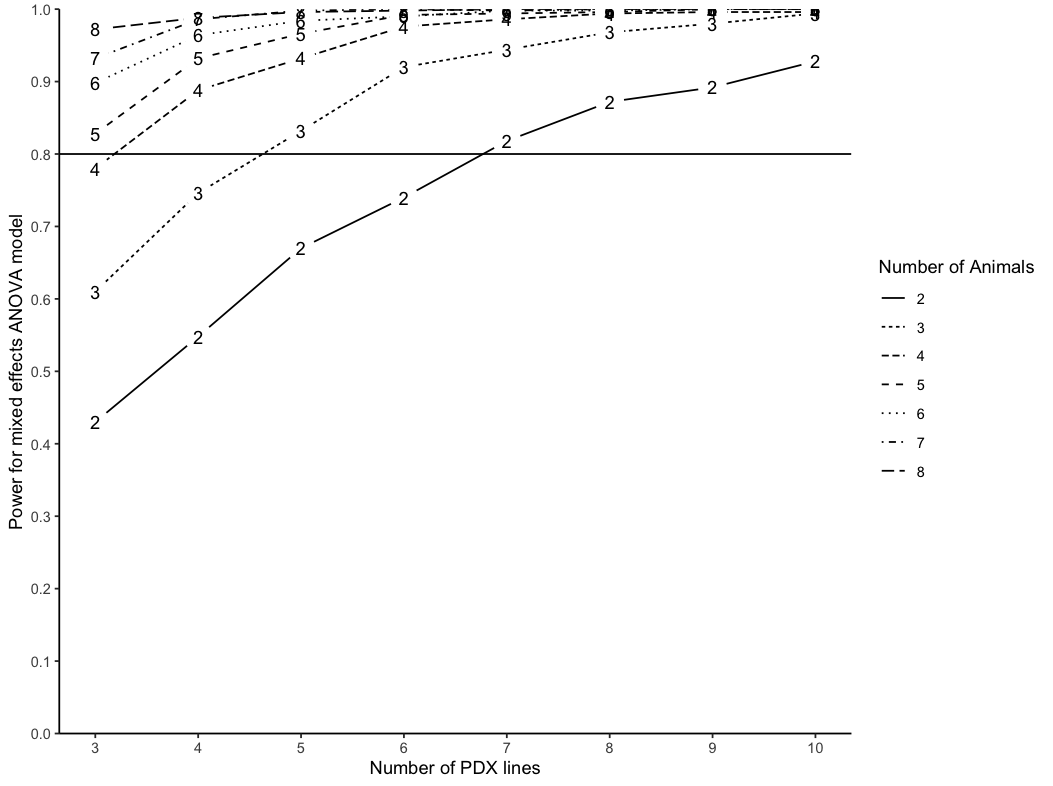}
    \caption{Power curve of the illustrating example for the mixed effects ANOVA model when there is no preliminary data available.}
    \label{fig:FigPower2nodat}
\end{figure}

One can also visualize the results from the above table, as shown in Figure \ref{fig:FigPower2nodat}, by calling the function \texttt{plotpower} in our package \textbf{PDXpower}. 
\begin{example}
### Visulaize the above table
plotpower(PowTab, ylim = c(0, 1))
\end{example}

Likewise in Section \ref{sec:pilotdata}, the above table displays the power for each combination of $n$ and $m$, calculated as the proportion of rejecting the null hypothesis $\beta=0$ based on Monte Carlo samples generated from a ANOVA mixed effect model (\ref{simANOVA}). It is observed from Figure \ref{fig:FigPower2nodat} that to achieve approximately 80\% power at a 5\% significance level, the minimal sample size of PDX lines and animals per arm per line per treatment group can be any of the following $(n,m)$ combinations: $\{(3, 5), (5, 3), (7, 2)\}$.

\subsection{Power analysis for the mixed crossed/nested design with censored data}
If some of the animals are alive (right-censored) at the end of an experiment, then Cox's frailty model (\ref{Cox}) should be considered to account for the censoring information. 
\subsubsection{Power analysis based on preliminary data}
\begin{figure}
    \centering
    \includegraphics[width=9cm]{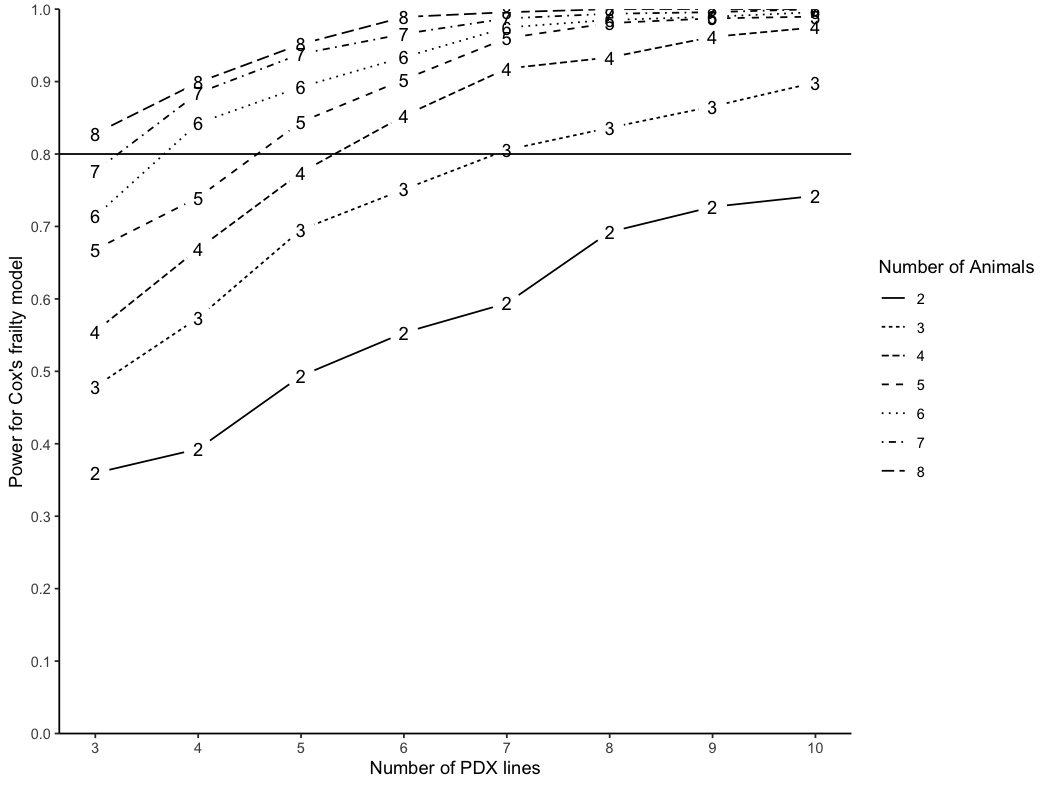}
    \caption{Power curve of the illustrating example for Cox's frailty model with preliminary data.}
    \label{fig:Figfrailty}
\end{figure}

%In certain pre-clinical studies, researchers may possess preliminary data from initial or related experimental explorations before embarking on a larger study. Priori information, as discussed in Section \ref{sec:COXpower},  can be derived from the preliminary data, aiding  simulation-based power analysis to determine the required number of PDX lines and animals per line per treatment group.
Similar to Section \ref{sec:pilotdata}, we have generated a censored preliminary dataset named \verb'animals2' through a simulation, which is stored in our package. This \verb'animals2' dataset comprises 18 animals and includes four columns: ID (PDX line ID number), Y (survival time of each mouse), Tx (treatment indicator with 1 for treatment and 0 for control), and status (death status with 1 for dead and 0 for alive). A screenshot of this dataset, produced as an output of the following R code, is depicted below. 
\begin{example}
data(animals2)
animals2

   ID         Y Tx status
1   1  5.097348  0      1
2   1  1.446492  0      0
3   1  4.230326  0      1
4   1 11.769290  1      1
5   1 10.791371  1      1
6   1  4.127112  1      0
7   2  4.644014  0      1
8   2  7.199358  0      1
9   2  7.727166  0      1
10  2  2.248739  1      1
11  2 11.879063  1      1
12  2  9.307388  1      1
13  3  1.726963  0      1
14  3 12.821633  0      1
15  3  2.918010  0      1
16  3  5.162990  1      1
17  3  6.672597  1      1
18  3 11.787588  1      0
\end{example}
After loading \verb'animals2' in Rstudio, a call to the function \texttt{PowFrailtyDat} derives the required priori information by fitting Cox's frailty model (\ref{Cox}) on \verb'animals2', and subsequently performs simulation-based power analysis for Cox's frailty model (\ref{Cox}) using the derived 
priori information for various sample size combinations.
\begin{example}
### Power analysis by assuming the time to event 
### is Weibull-distributed
PowTab <- PowFrailtyDat(data = animals2, 
formula = Surv(Y,status) ~ Tx + cluster(ID),
n = 3:10, m = 2:8, 
Ct = 12, censor = TRUE, 
sim = 500)
\end{example}
The inputs for the function PowFrailtyDat include \texttt{data}, \texttt{formula}, the number of PDX lines \texttt{n}, the number of animals \texttt{m}, \texttt{Ct} duration of follow-up, and \texttt{censor} whether type I censoring is considered (\texttt{Ct} is then specified if \texttt{TRUE}), and the number of Monte Carlo replicates \texttt{sim}. The output, saved in PowTab, is shown below.
{\small
\begin{example}
Parameter estimates based on the pilot data:
Scale parameter (lambda): 0.0154 
Shape parameter (nu): 2.1722 
Treatment effect (beta): -0.8794 
Variance of random effect (tau2): 0.0422 

Monte Carlo power estimate, calculated as the
proportion of instances where the null hypothesis
H_0: beta = 0 is rejected (n = number of PDX lines,
m = number of animals per arm per PDX line,
N = total number of animals for a given combination
of n and m, 
Censoring Rate = average censoring rate across 500 
Monte Carlo samples):
    n m   N Power (%) for Cox's frailty Censoring Rate
1   3 2  12                     35.96          13.88
2   3 3  18                     47.89          14.01
3   3 4  24                     55.51          14.07
4   3 5  30                     66.82          14.01
5   3 6  36                     71.40          14.54
6   3 7  42                     77.63          14.16
7   3 8  48                     82.81          14.09
8   4 2  16                     39.29          13.99
9   4 3  24                     57.38          14.10
10  4 4  32                     66.89          14.15
11  4 5  40                     74.00          13.76
12  4 6  48                     84.26          14.41
13  4 7  56                     88.42          14.28
14  4 8  64                     89.89          14.31
15  5 2  20                     49.37          14.00
16  5 3  30                     69.56          14.13
17  5 4  40                     77.46          14.26
18  5 5  50                     84.43          14.13
19  5 6  60                     89.26          14.47
20  5 7  70                     93.83          14.11
21  5 8  80                     95.18          14.13
22  6 2  24                     55.33          14.08
23  6 3  36                     75.12          14.14
24  6 4  48                     85.27          14.54
25  6 5  60                     90.27          14.12
26  6 6  72                     93.36          14.38
27  6 7  84                     96.58          14.04
28  6 8  96                     98.92          14.16
29  7 2  28                     59.40          14.20
30  7 3  42                     80.57          14.05
31  7 4  56                     91.80          14.55
32  7 5  70                     95.99          14.11
33  7 6  84                     97.47          14.21
34  7 7  98                     98.72          14.11
35  7 8 112                     99.57          14.18
36  8 2  32                     69.23          14.11
37  8 3  48                     83.63          14.01
38  8 4  64                     93.33          14.48
39  8 5  80                     98.08          13.95
40  8 6  96                     98.51          14.32
41  8 7 112                     99.36          14.09
42  8 8 128                    100.00          14.23
43  9 2  36                     72.70          14.11
44  9 3  54                     86.57          13.96
45  9 4  72                     96.17          14.35
46  9 5  90                     98.72          13.94
47  9 6 108                     98.95          14.30
48  9 7 126                     99.58          14.15
49  9 8 144                    100.00          14.24
50 10 2  40                     74.23          14.28
51 10 3  60                     89.88          13.95
52 10 4  80                     97.48          14.43
53 10 5 100                     98.97          13.98
54 10 6 120                     99.58          14.37
55 10 7 140                    100.00          14.16
56 10 8 160                    100.00          14.24
\end{example}
}
Similarly, one may generate a graphical representation of the estimated power listed in the above table, as shown
in Figure \ref{fig:Figfrailty}, by calling the function \texttt{plotpower}:
\begin{example}
### Visualize the above table 
plotpower(PowTab[[5]], ylim = c(0, 1))
\end{example}
It is observed from Figure \ref{fig:Figfrailty} that to achieve approximately 80\% power at a 5\% significance level, the minimal sample size of PDX lines and animals per arm per line per treatment group can be any of the following $(n,m)$ combinations: $\{(3, 8), (4, 6), (5, 5), (6, 4), (7, 3)\}$.

\subsubsection{Power analysis without preliminary data}
Similar to Section \ref{sec:nopilotdata}, when preliminary data on the treatment under investigation are not available, Table \ref{tab:tab2} provides a summary of the required information for power calculations using Cox's frailty model for the mixed crossed/nested PDX experimental design if one assumes the time-to-event outcome follows a Weibull distribution. 

%then the function \texttt{PowFrailty} may be called to generate Weibull survival Monte Carlo samples (see equation (\ref{simCox})) based on the median survival of both groups and allows for Monte Carlo simulation-based power calculation over sample size combinations. 
The power calculations in the following example assume a constant hazard ($\nu=1$, set as default), endpoint of the study period (\texttt{Ct}=12), and an approximate variance of the log survival across PDX lines ($\tau^2=0.1$, set as default, i.e., 0.31 unit of standard deviation in treatment effect $\beta$ across PDX lines).

\begin{table}
\resizebox{9cm}{!} 
{ 
\begin{tabular}{@{}llll}
\hline \hline
Parameter & Interpretation   & Values\\
\hline
{\bf Cox's frailty} & \\
\\[-.1in]
ctl.med.surv & Median survival in the control arm & Any positive number\\
tx.med.surv &Median survival in the treatment arm & Any positive number\\[.1in]
 $\nu$     & Baseline Weibull shape parameter     & $\nu = 1$ constant hazard, default\\
          &                                   & $\nu > 1$ increasing hazard \\
           &                                  & $\nu < 1$ decreasing hazard\\
\\
$\tau^2$  & Heterogeneity of PDOX lines.  & $\tau^2 = 0$: no heterogeneity\\
          & $-$ Set to assumed variance in  & $\tau^2 > 0$: heterogeneity\\
          & log survival between PDOX lines & set $\tau^2 = 0.1$ as default\\
\\[-.1in]
 \hline
 \hline
\end{tabular}
}
\caption{Parameter value elicitation for Weibull distributed data for given number of PDX lines $n$ and animals per cell line $m$.}
\label{tab:tab2}
\end{table}

\begin{figure}
    \centering
    \includegraphics[width=9cm]{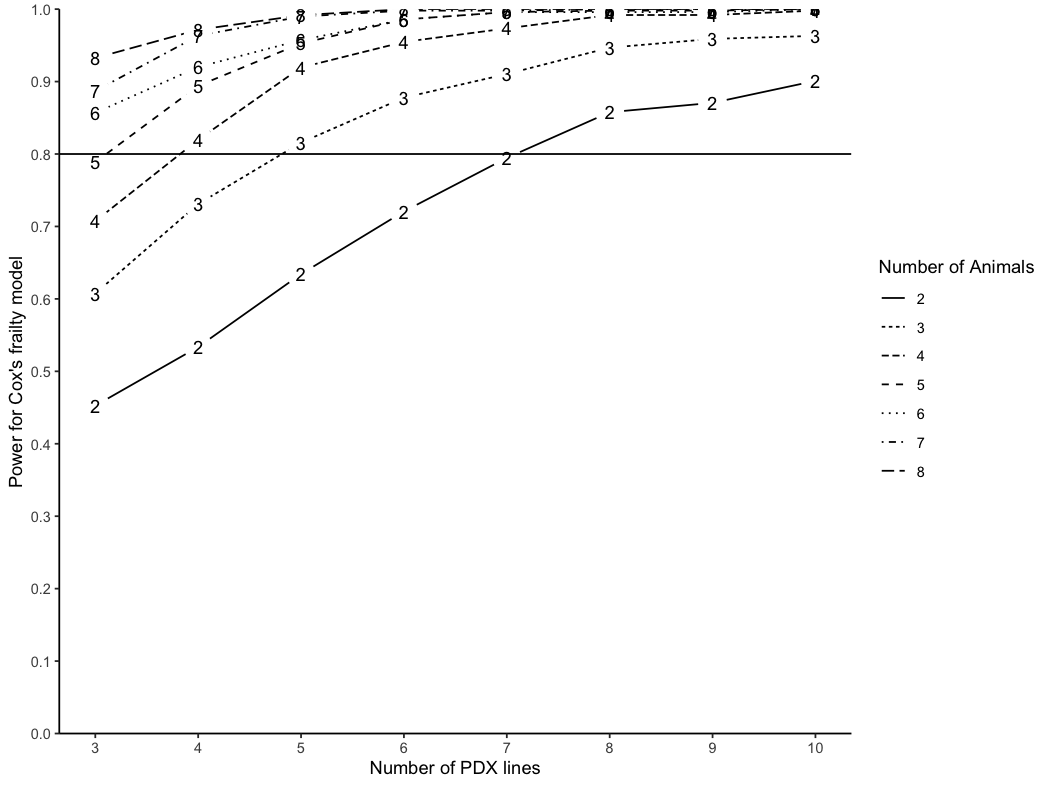}
    \caption{Power curve of the illustrating example for Cox's frailty model when there is no preliminary data available.}
    \label{fig:Figfrailtynodat}
\end{figure}

\begin{example}
### Power analysis by specifying the median survival
### of control and treatment group and assuming 
### the time-to-event is Weibull-distributed
PowTab <- PowFrailty(ctl.med.surv = 2.4, 
tx.med.surv = 7.2, nu = 1, tau2 = 0.1, sim = 500,
censor = TRUE, Ct = 12,
n = 3:10, m = 2:8)
\end{example}
The output, saved in PowTab, is shown below.
{\small
\begin{example}  
Treatment effect (beta): -1.098612 
Scale parameter (lambda): 0.2888113 
Shape parameter (nu): 1 
Variance of random effect (tau2): 0.1 

Monte Carlo power estimate, calculated as the
proportion of instances where the null hypothesis
H_0: beta = 0 is rejected (n = number of PDX lines,
m = number of animals per arm per PDX line,
N = total number of animals for a given combination
of n and m, 
Censoring Rate = average censoring rate across 500 
Monte Carlo samples):
    n m   N Power (%) for Cox's frailty Censoring Rate
1   3 2  12                     45.26          17.72
2   3 3  18                     60.74          17.93
3   3 4  24                     70.73          18.13
4   3 5  30                     78.88          17.63
5   3 6  36                     85.69          18.37
6   3 7  42                     88.71          17.84
7   3 8  48                     93.25          17.62
8   4 2  16                     53.36          17.82
9   4 3  24                     73.10          17.89
10  4 4  32                     81.97          18.22
11  4 5  40                     89.41          17.44
12  4 6  48                     92.01          18.17
13  4 7  56                     96.33          17.98
14  4 8  64                     97.15          17.93
15  5 2  20                     63.43          17.62
16  5 3  30                     81.57          17.91
17  5 4  40                     91.94          18.28
18  5 5  50                     95.27          17.84
19  5 6  60                     95.73          18.34
20  5 7  70                     98.98          17.80
21  5 8  80                     99.17          17.77
22  6 2  24                     72.07          17.72
23  6 3  36                     87.78          17.85
24  6 4  48                     95.47          18.47
25  6 5  60                     98.57          17.90
26  6 6  72                     98.57          18.24
27  6 7  84                     99.79          17.71
28  6 8  96                    100.00          17.83
29  7 2  28                     79.46          17.85
30  7 3  42                     91.04          17.80
31  7 4  56                     97.34          18.49
32  7 5  70                     99.59          17.89
33  7 6  84                     99.59          18.07
34  7 7  98                     99.79          17.82
35  7 8 112                    100.00          17.89
36  8 2  32                     85.77          17.82
37  8 3  48                     94.68          17.78
38  8 4  64                     99.19          18.41
39  8 5  80                     99.80          17.75
40  8 6  96                     99.59          18.18
41  8 7 112                    100.00          17.83
42  8 8 128                    100.00          18.01
43  9 2  36                     87.06          17.89
44  9 3  54                     95.89          17.78
45  9 4  72                     99.19          18.25
46  9 5  90                     99.80          17.79
47  9 6 108                     99.59          18.16
48  9 7 126                    100.00          17.94
49  9 8 144                    100.00          17.98
50 10 2  40                     90.10          18.18
51 10 3  60                     96.31          17.73
52 10 4  80                     99.80          18.27
53 10 5 100                    100.00          17.85
54 10 6 120                    100.00          18.24
55 10 7 140                    100.00          17.91
56 10 8 160                    100.00          17.94
\end{example}
}
Similarly, one may generate a graphical representation of the estimated power listed in the above table, as shown
in Figure \ref{fig:Figfrailtynodat}, by calling the function \texttt{plotpower}:
\begin{example}
### Visualize the above table 
plotpower(PowTab, ylim = c(0, 1))
\end{example}
It is observed from Figure \ref{fig:Figfrailtynodat} that to achieve approximately 80\% power at a 5\% significance level, the minimal sample size of PDX lines and animals per arm per line per treatment group can be any of the following $(n,m)$ combinations: $\{(3, 5), (4, 4), (5, 3), (7, 2)\}$.

\section{Discussion}
\label{sec:dis}
%Pre-clinical studies relying on PDX are essential tools for the development of novel therapies. The use of multiple PDX lines, and multiple animal model per PDX line,  allows for explorations and experimental adjustment to the effect of heterogeneous genetic background of patient tumors \citep{cassidy2015maintaining,giannini2005patient,goodspeed2016tumor}. 
We have introduced an R package, `\textbf{PDXpower}',  designed for simultaneously determining the required number of both PDX lines and animals per line per treatment group for PDX experiments under the mixed crossed/nested design for either uncensored or  right-censored time-to-event outcomes. We have also provided step-by-step tutorials for its utilization, accommodating scenarios with or without preliminary data. For a right-censored outcome, our package is tailored to type I censoring because it is typical in a PDX experiment for all animals to start treatment at the same time and subsequently be subject to right-censoring by the same administrative censoring at the end of follow-up.

Our power calculation strategy is simulation-based, involving fitting either the mixed effects ANOVA model for uncensored outcomes or Cox's frailty model with a normal random effect for right-censored outcomes over a large number of Monte Carlo samples. %It can be computationally intensive when the number of Monte Carlo samples is large, say 2,000, even though 
We have implemented a parallel computing strategy in our package to speed up the computation. Based on our experience, our package is fast for the mixed effects ANOVA model for uncensored outcomes as it took around 0.5 minutes to run 500 Monte Carlo samples to generate Figure \ref{fig:FigPower2} 
%considering all possible combinations of number of PDX lines and animals per line, 
on a MacBook Pro with 8-Core M1 Pro and 16GB RAM running MacOS.
%(see the examples in Section \ref{sec:example}). 
However, it took about 20 minutes to generate the power curves in Figure \ref{fig:Figfrailty} based on Cox's frailty model for right-censored outcomes. One could obtain results more quickly by considering a coarser grid for $n$ and $m$, resulting in fewer combinations. To enhance the computational efficiency of power calculation for censored data, one could also explore alternate working models, such as different frailty models like the gamma frailty model \citep{chen2014sample} or standard fixed-effects Cox models treating PDX as fixed effects. Further investigations into these models are warranted. 

Our \textbf{PDXpower} package requires users to have some basic working knowledge of R \citep{R}. We also plan to build interactive web apps based on \textbf{PDXpower} using Rshiny \citep{Rshiny}, which will enable users with no R knowledge to perform power analysis for PDX experiments under the mixed crossed/nested design.

\section*{Software}
An R package, \textbf{PDXpower}, designed for conducting simulation-based power analysis in this paper, is publicly available at The Comprehensive R Archive Network \url{https://CRAN.R-project.org/package=PDXpower}.

\section*{Acknowledgement}
This research was partially supported by National Institutes of Health (P50CA211015, P30 CA-16042, R01NS121319, P50 CA092131, and P01CA236585, GL, P30 CA-033572, JP), and the National Center for Advancing Translational Sciences (UL1-TR-001420, GL). 

\bibliographystyle{apalike}
\bibliography{reference}

%% Table %%
%%%%%%%%%%%%%%%%%%%%%%
%\begin{table}
%\caption{}\label{}
%\end{table}

%% Figure %%
%%%%%%%%%%%%%%%%%%%%%%
%\begin{figure}[t]
%\includegraphics{}
%\caption{}\label{}
%\end{figure}

%% Theorem %%
%%%%%%%%%%%%%%%%%%%%%%
%%\begin{thm}[]\label{} Theorem
%%\end{thm}

%% Proof %%
%%%%%%%%%%%%%%%%%%%%%%
%\begin{proof}
%\end{proof}

%% Acknowledgements %%
%%%%%%%%%%%%%%%%%%%%%%
%\begin{acknowledgement}%[title={Acknowledgments}]
%\end{acknowledgement}

%% Funding %%
%%%%%%%%%%%%%%%%%%%%%%
%\begin{funding}
%\end{funding}

%% Supplementary Material %%
%%%%%%%%%%%%%%%%%%%%%%
%\begin{esm}
%\esmtitle{Supplementary Material}
%\esmdescription{}
%\esmfilename{}
%\end{esm}

%% Appendices %%
%%%%%%%%%%%%%%%%%%%%%%
%\begin{appendix}
%\section{Appendix section}
%\end{appendix}

%% Bibliography %%
%%%%%%%%%%%%%%%%%%%%%%
%\bibliographystyle{nessart-number}
%\bibliography{biblio}

%% restore discarded spaces in the last balanced column
%\atColsBreak{\pagediscards}

\end{document}